\documentclass[12pt]{article}
\input psfig.tex
\usepackage[dvips]{graphicx}
\usepackage[hang]{subfigure}
\newcommand{\bfi}{\begin{figure}}
\newcommand{\efi}{\end{figure}}
\newcommand{\subf}{\subfigure}
\newcommand{\bmi}{\begin{minipage}}
\newcommand{\emi}{\end{minipage}}

\begin{document}
\centerline{\large\bf Universality and Scale Invariance in
 Hourly Rainfall}

\bigskip
\centerline{\large \bf Pankaj Jain, Suman Jain and Gauher Shaheen}

\bigskip
\centerline{Physics Department, I.I.T. Kanpur, India, 208016}

\bigskip
\noindent
{\bf Abstract:}
We show that the hourly rainfall rate distribution can be described
by a simple power law to a good approximation. We show that the
exponent of the distribution in tropics is universal and is equal to 
$1.13\pm 0.11$. At higher latitudes the exponent increases and is found
to lie in the range 1.3-1.6. 

\vskip 0.3in
It is well known that daily rainfall over a localized region displays 
an almost chaotic behavior \cite{chaos}. 
This implies that it is very difficult to make even short term prediction 
of daily rainfall. 
The distribution of daily rainfall amount over a localized region is 
also well known to have a very large tail which implies that there is very 
high probability to have very large rainfall. The distribution shows 
considerable deviation from the normal distribution.

In the present paper we analyze the distribution of hourly rainfall amount.
We are particularly 
interested in determining whether we can identify some universal features 
in the rainfall distribution which are observed at all spatial locations. 
The nature of distribution can also reveal some underlying features of 
the dynamical system. For example, it has been argued that dissipative 
systems with large number of variable tend to reach a critical state under 
the influence of an external perturbation. In the critical state the 
system displays an almost chaotic behavior with fluctuations displaying 
scale invariant power law distribution. This has been demonstrated 
numerically by postulating simplified models of the behavior of sand 
piles, earthquakes etc \cite{SOC1,SOC2}. 
The mechanism by which the dynamical systems 
achieve the critical state under the influence of external perturbation 
is called the self organized criticality \cite{SOC1,SOC2}. 
In the present paper we shall
determine if the rainfall rate also displays this behavior by 
studying its distribution. Some preliminary results of this study 
have been presented in Ref. \cite{Jain}. 

The data is taken from the http://daac.gsfc.nasa.gov web site 
and consists of SSM/I satellite data. The data consists of rainfall
rate in units of mm/hour with the minimum rate equal to 0.1 mm/hour. 
It is available over the entire globe on a grid size of half a degree 
both in latitude and longitude. 
 We extracted the data
files from this website for the year 1997 for three different time
periods. At different times the data is available at different locations.
We randomly selected a total of 97 different locations on the globe
out of the extracted data. 
In order that a reasonable
amount of data is available for a given region, we group the rainfall data in
intervals of 5 degrees both in latitude and longitude. The typical distribution
for several different regions are shown in Fig. \ref{Tropics} and
\ref{High_Lat}. 
We represent the power law distribution as 
\begin{equation}
f(x) = a/x^b
\end{equation}
where $x$ is equal to the hourly rainfall amount in mm and $a$ and $b$ are
parameters. 
If the underlying dynamical system is in a critical state then we expect
that certain observables in the system will follow a power law distribution. 
The rainfall amount per hour represents an interesting
physical observable which we study in this paper.

We point out that while testing for a power law behaviour
we are not testing the hypothesis that the distribution is a power
law over the entire interval of rainfall per hour. The power law
is expected to be valid only in some intermediate range of rainfall. 
At low as well as large rainfall we expect that the distribution
will be distorted in comparison to a pure power behavior. At very large
rainfall, for example, we expect the distribution to decay very 
rapidly since rainfall amounts larger than a certain value are 
physically not possible. We, therefore, cut off the distribution
at some large value as well as at a low value of the rainfall. At the
low rainfall value we make the distribution constant below a certain
value $x_{\rm min}$, which is treated as an additional parameter of the
distribution. At the large rainfall end we put a cut on the data so as
to exclude all data with $x> x_{\rm max} =10$ mm/hr. 
The number of data points in the
excluded region are typically found to be very small.   
 The distribution is set equal to zero beyond $x=x_{\rm max}$.
The power distribution, therefore, contains two adjustable
parameters $b$ and $x_{\rm min}$. The parameter $a$ is
obtained by normalization of the distribution. 

We compare the fit obtained by the power law distribution with an
alternative exponential distribution
\begin{equation}
g(x) = \alpha \exp(-\beta x)
\end{equation}
In analogy with the power distribution, $g(x)$ is also set equal
to zero for $x>x_{\rm max}$. Hence the normalization $\alpha$
for this distribution is equal to $\beta/(1-\beta x_{\rm max})$.  
We point out that we are not interested in
finding the best possible distribution that describes the data. We
are only interesting in demonstrating that a power law provides a
good description of the data over a wide range of rainfall amounts.
Hence we have not made extensive comparisons of various distributions
and compare the power law only to a simple exponential fit. 

\begin{figure}
\psfig{file=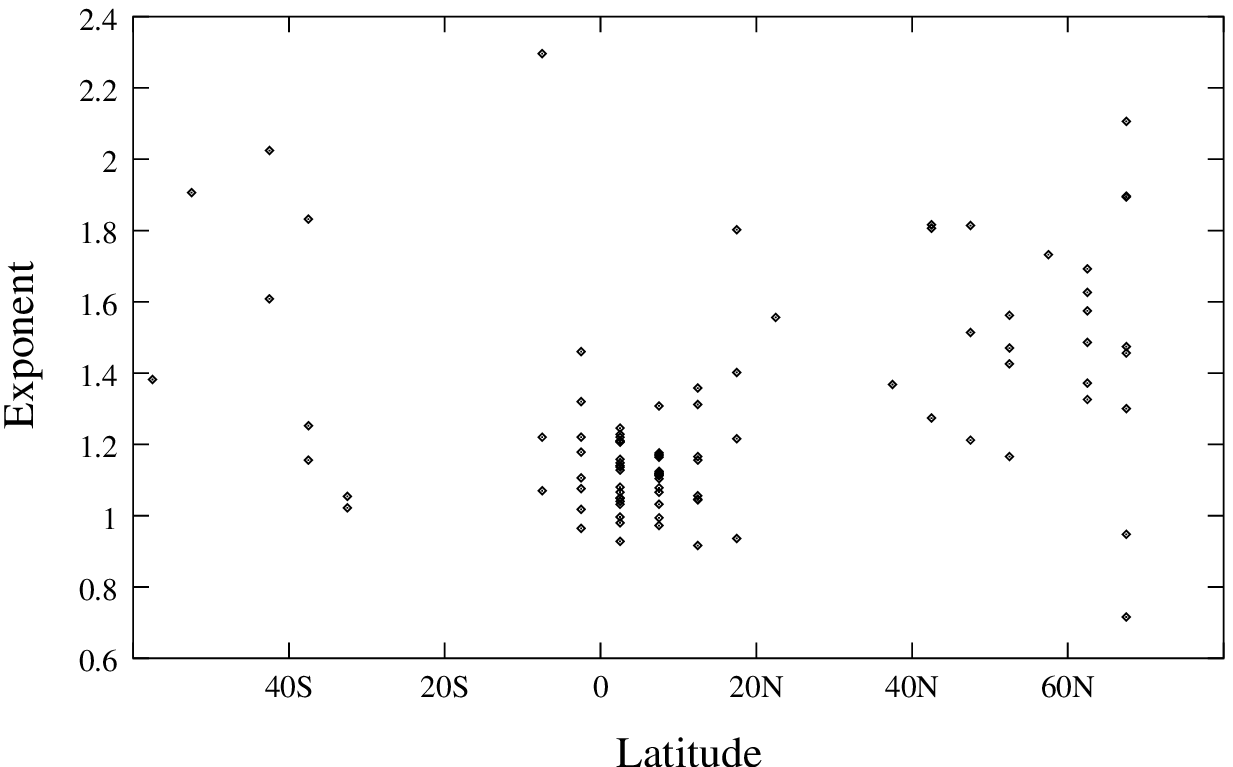}
\caption{Scatter plot of the exponent of the power distribution 
fits as a function of the latitude.}
\label{Lat_Slope}
\end{figure}

\begin{figure}
\psfig{file=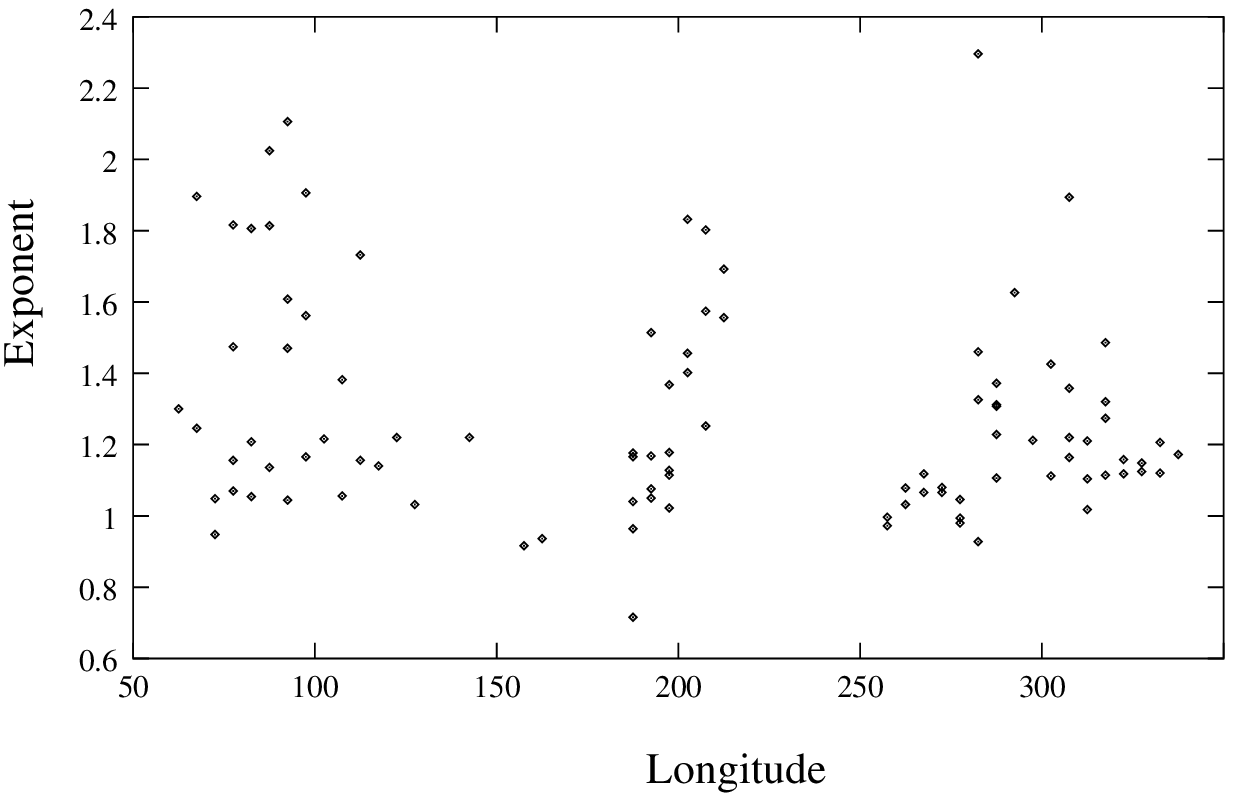}
\caption{Scatter plot of the exponent of the power distribution 
fits as a function of the longitude.}
\label{Long_Slope}
\end{figure}

In Fig. \ref{Tropics} and \ref{High_Lat} 
we show the fits in several representative cases.
The figure shows the power as well as the exponential fit to the 
rainfall distribution. It is clear from these plots that a 
power distribution fits the data reasonably well. It
can also be seen that the fit is much better at lower 
latitudes in comparison to the higher latitudes.
The log likelihood difference, defined to be the difference of the 
log likelihood for the power fit and the exponential fit, 
is generally found to lie in the range 50 to 150 in tropics
and between 10 to 30 at the higher latitudes.
Hence the power distribution gives an overall much better description of
the data in comparison to an exponential distribution.
In Fig. \ref{Lat_Slope} and \ref{Long_Slope} 
we show a scatter plot of the exponent of the power
distribution as a function of the latitude and longitude respectively. 
The error in the exponent values ranges from 0.02 to 0.1 in most cases.
Only rarely does the error exceeds 0.1. The precise value depends
on the number of data points obtained in each region which typically
range from 100 to 2000. We also point out that the uneven distribution
of the data points as a function of latitude and longitude is caused
by our selection of times for which the data is extracted. However
this does not have any influence on our results or conclusions since
the data is spread over the entire globe.

We clearly see that the values of the exponent $b$ are different at
lower latitudes in comparison to higher latitudes. The exponent close
to equator is found to be close to unity whereas at higher latitudes
the exponent is much larger. The exponent does not show very
significant dependence 
on the longitude. Only a very marginal relationship is found after
eliminating one outlier with a very large value of
the exponent.
In order to quantify the dependence of the exponent on the
latitude we evaluate the correlation between the absolute value of
the latitude and the exponent. For the entire set of 97 data
points the correlation 
$\rho=0.54$. If we delete two outliers the correlation goes up to
$\rho=0.608$. The probability $p$ that we can get this correlation from
a random sample is very small. We find that for the entire set
$p=10^{-5}$\% and after eliminating the two outliers $p= 3\times 10^{-7}$\%.  

The fact that at lower latitude the exponent values are clustered around 
unity irrespective of longitude is a clear indication of a 
universal behaviour in the rainfall distributions in the tropics. By putting a
cut on the latitude in order to select only the region that
lies between 20 N and 20 S we find that the mean value of the exponent 
$b$, defined in Eq. 1,
is given by,
\begin{equation}
b = 1.13\pm 0.11
\label{Mean_Slope}
\end{equation}
The total number of data points contained within the latitude
20 N and 20 S are equal to 61. Two of these points were found
to give an anomalously large values of the exponent compared to 
the rest and were treated as
outliers. The mean value given in Eq. \ref{Mean_Slope} is obtained
after eliminating these outliers. The median value of the exponent is
equal to 1.12 and remains unchanged with or without the inclusion
of the outliers. 

As this work was near completion we became aware of Ref. \cite{Peters}
where the authors have analyzed high-resolution rainfall data
at Baltic coast Zingst 
in order to investigate its distribution. The authors
find that the distribution is well described by a power law
with exponent 1.36. Our results are in agreement with their findings
since at higher latitudes we also find similar exponents.  

In conclusion, we summarize the main results of the paper. We have
shown that the hourly rainfall distribution is well described by a 
scale invariant power law. This is particularly true in the tropical
region where the exponent is found to be $1.13\pm 0.11$ independent
of the longitude. As we go towards the higher latitudes the exponent
increases and generally is found to lie in a range 1.3-1.6.
The power dependence and its universal character in the tropics
indicates that the underlying dynamical system may be best 
describable in terms of self organized criticality.

\bigskip
\noindent
{\bf Acknowledgements:}
We thank Ralf Bennartz, Ashu Jain and Mahendra Verma
for useful discussions.
We also thank the Distributed Active Archive Center 
at the
Goddard Space Flight Center, Greenbelt, MD, 20771, for producing and
distributing the data which was used in this work.
Partial funding for this work is provided by DST.

\medskip
\noindent
All correspondence should be directed to Pankaj Jain (pkjain@iitk.ac.in).

\bfi
\subf{
\bmi[t]{3in}
\includegraphics[scale=0.55]{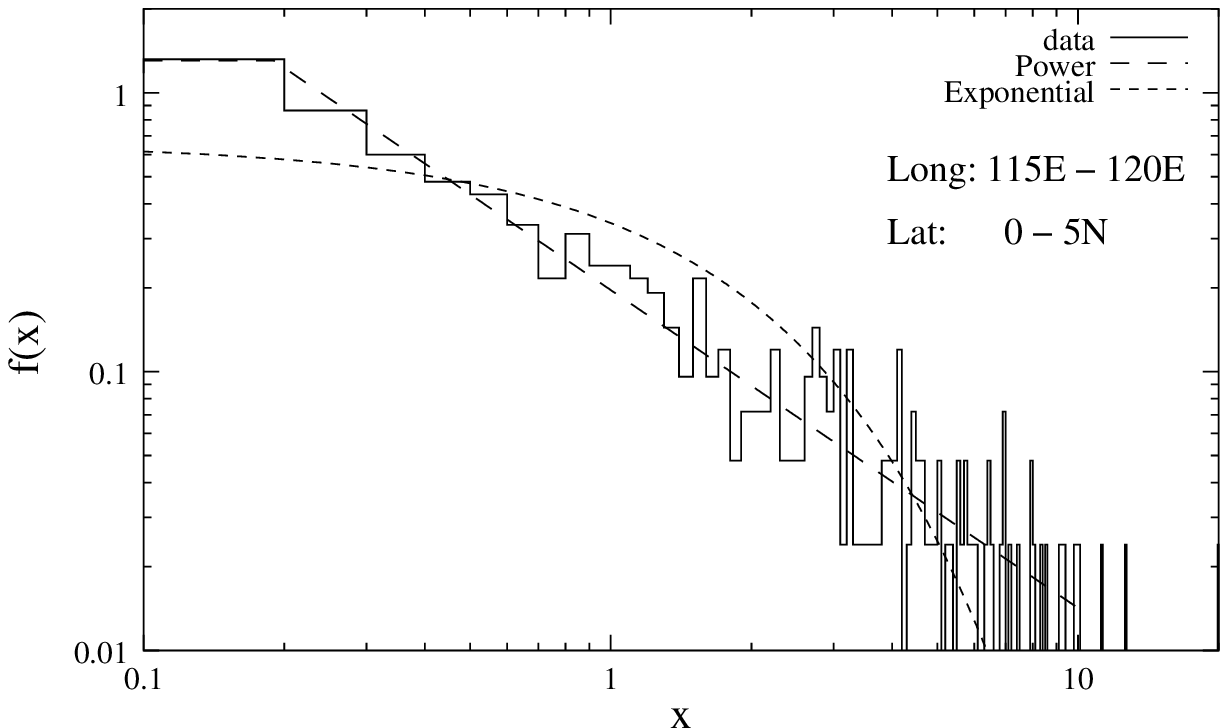}
\emi
}
\subf{
\bmi[t]{3in}
\includegraphics[scale=.55]{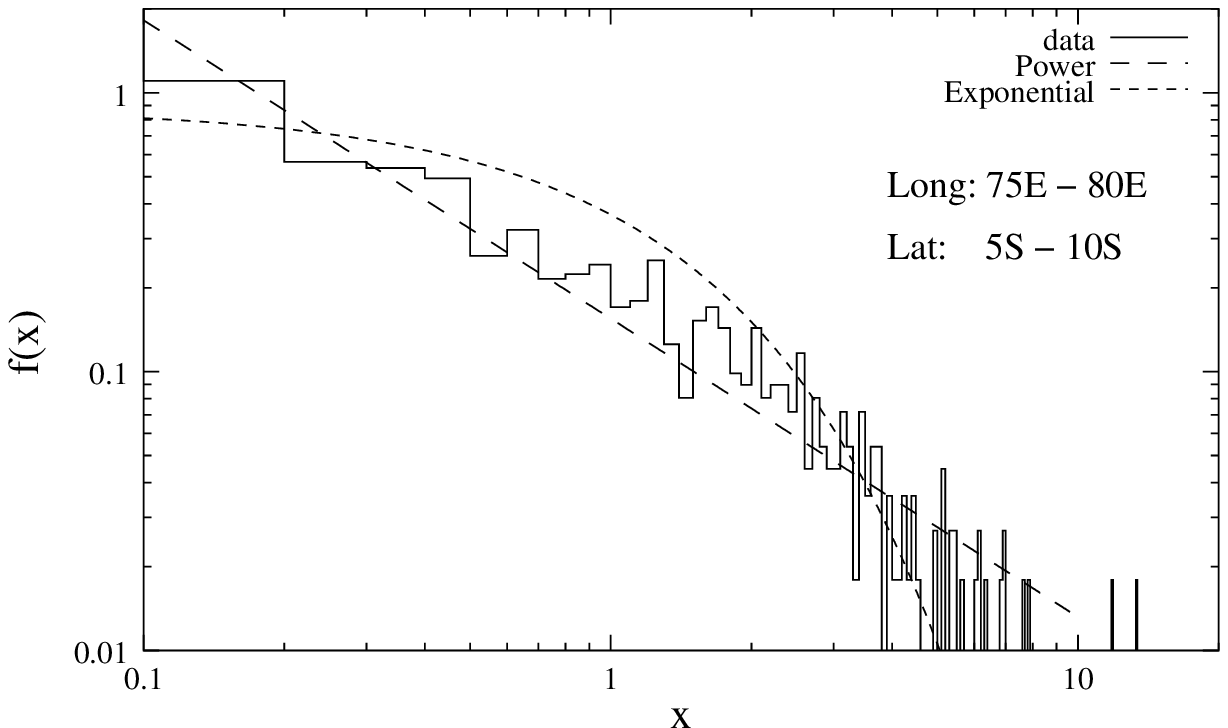}
\emi
}
\subf{
\bmi[t]{3in}
\includegraphics[scale=.55]{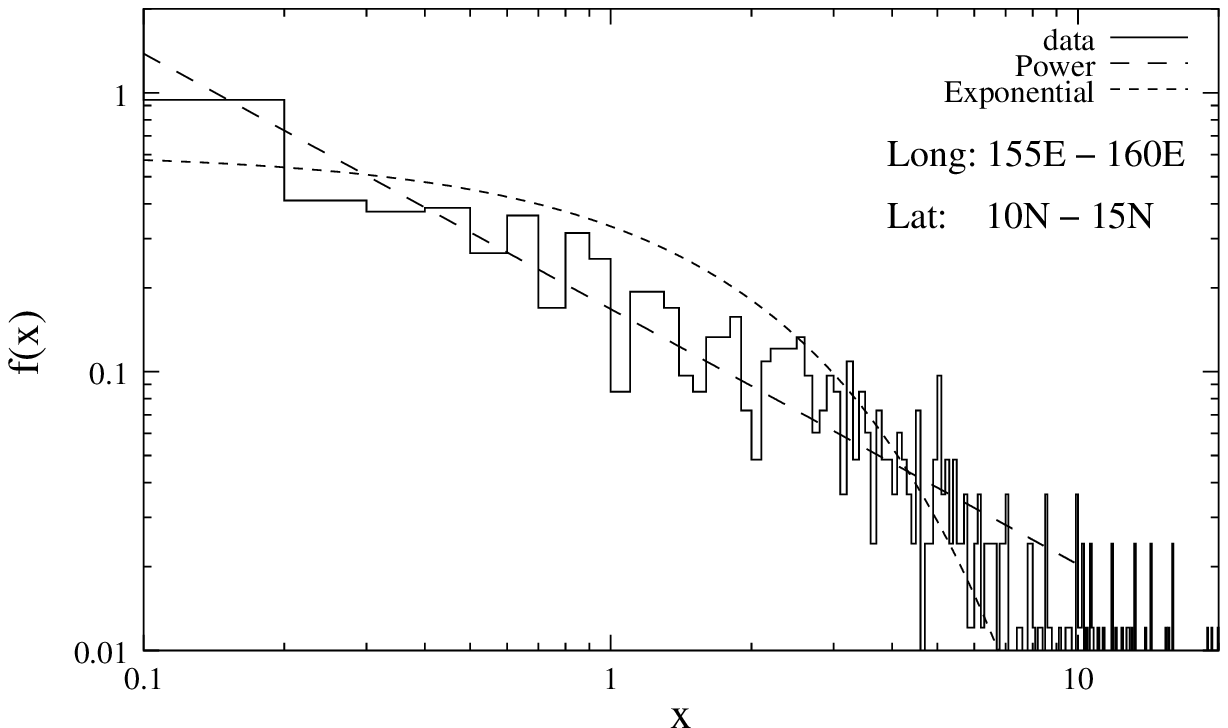}
\emi
}
\subf{
\bmi[t]{3in}
\includegraphics[scale=.55]{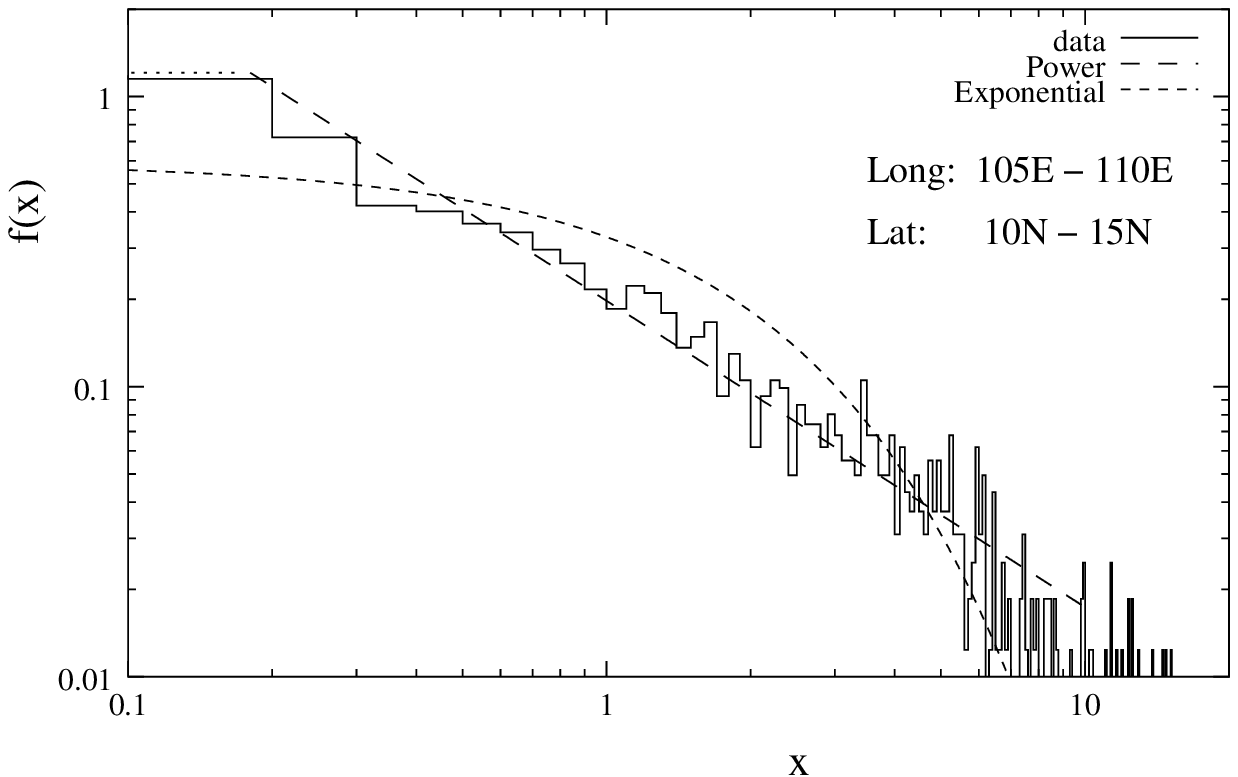}
\emi
}
\subf{
\bmi[t]{3in}
\includegraphics[scale=.55]{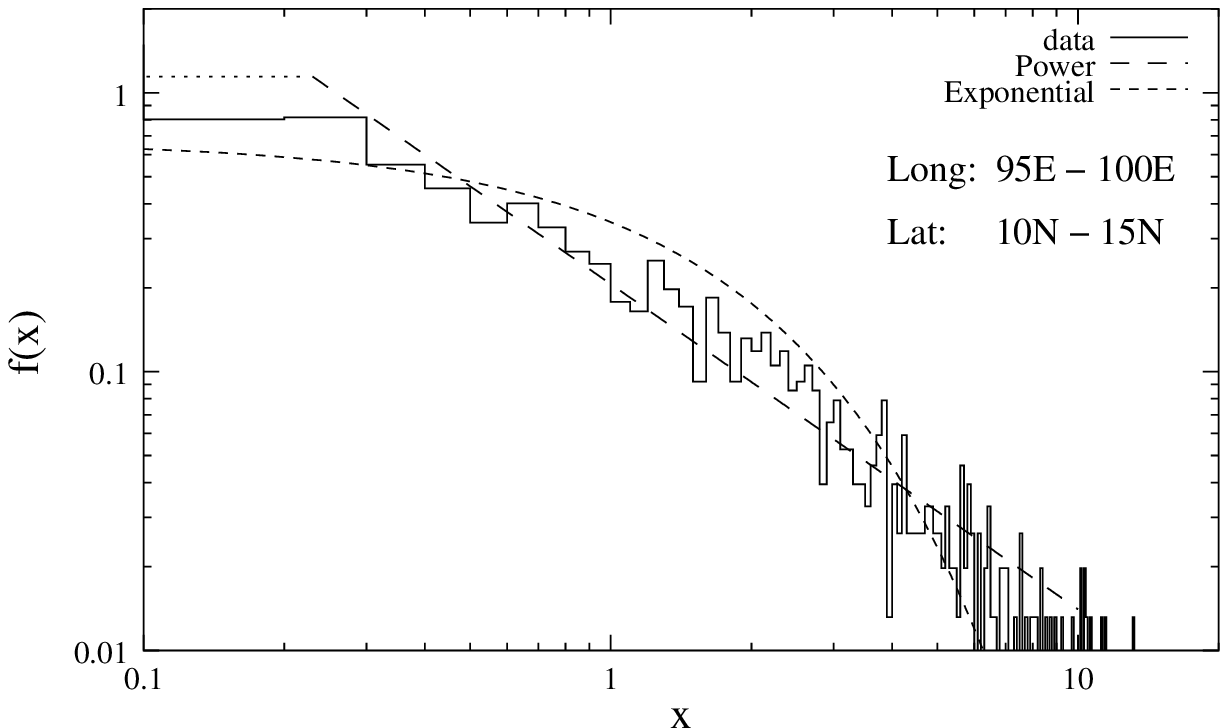}
\emi
}
\subf{
\bmi[t]{3in}
\includegraphics[scale=0.55]{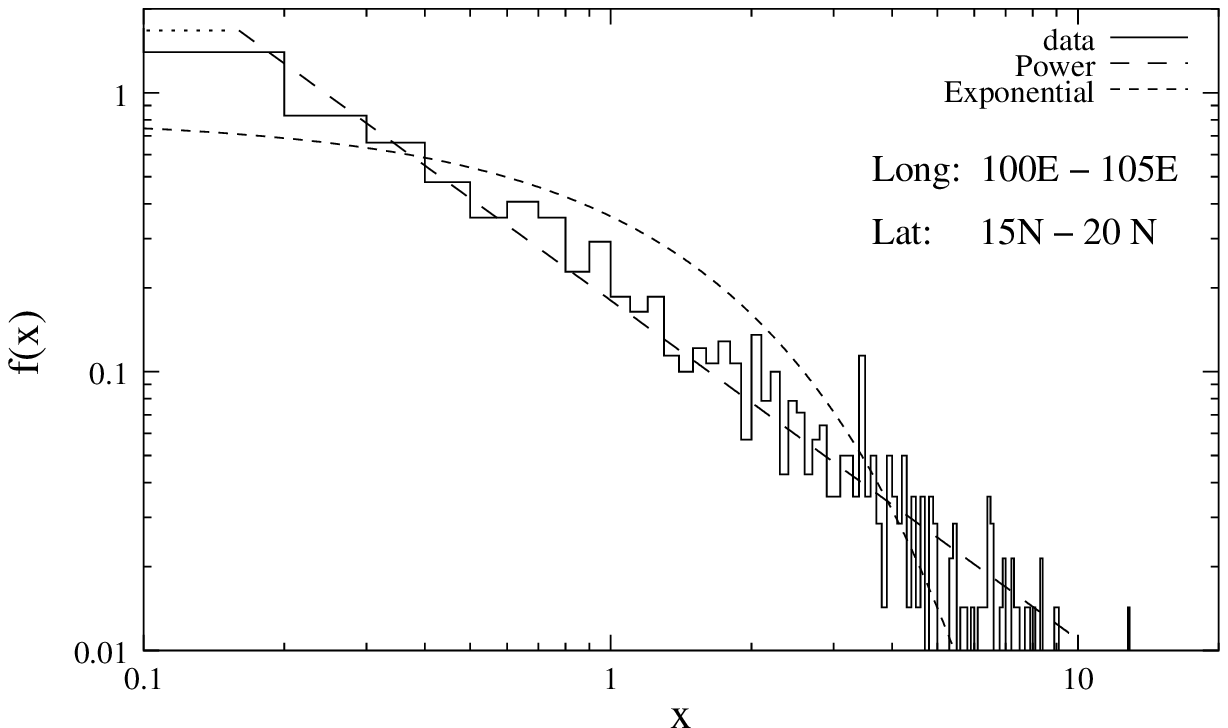}
\emi
}
\caption{Hourly rainfall distributions $f(x)$ of a sample of data sets in the 
the tropical region. The variable $x$ is the rainfall rate in mm/hour.  
The best fit power and exponential distributions
are also shown. The longitude and latitude range from where the 
data was taken is also indicated in each graph}
\label{Tropics}
\efi

\bfi
\subf{
\bmi[t]{3in}
\includegraphics[scale=0.55]{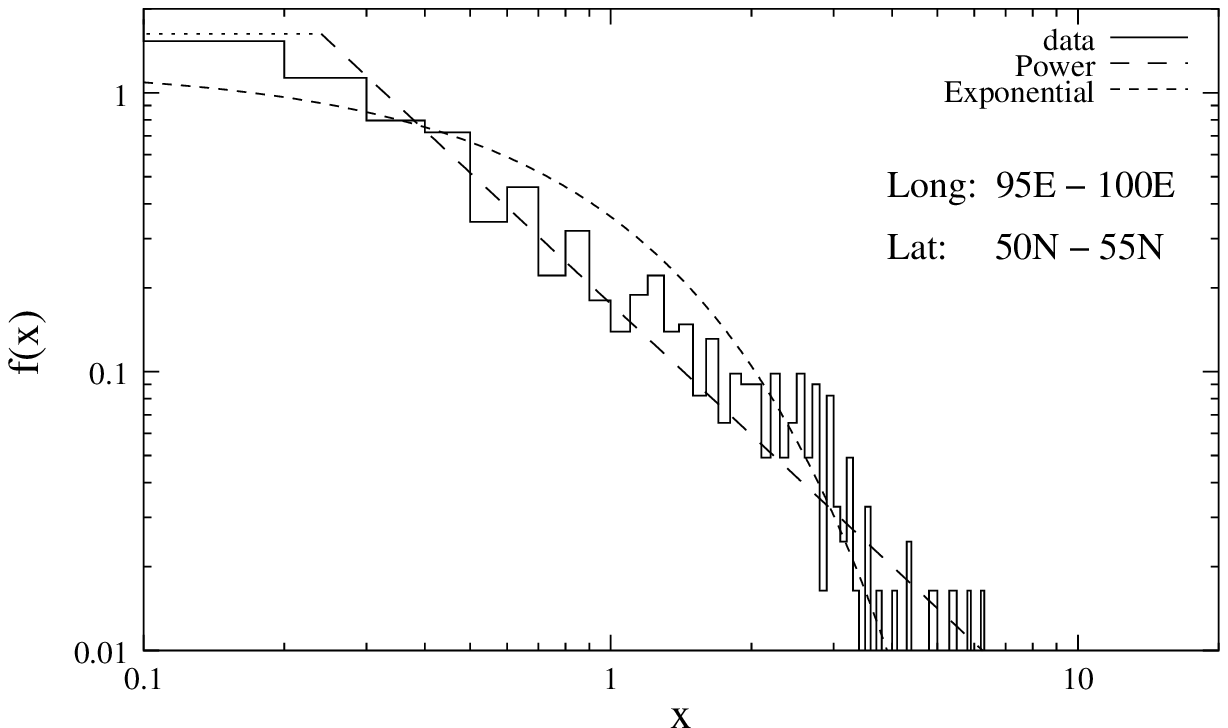}
\emi
}
\subf{
\bmi[t]{3in}
\includegraphics[scale=0.55]{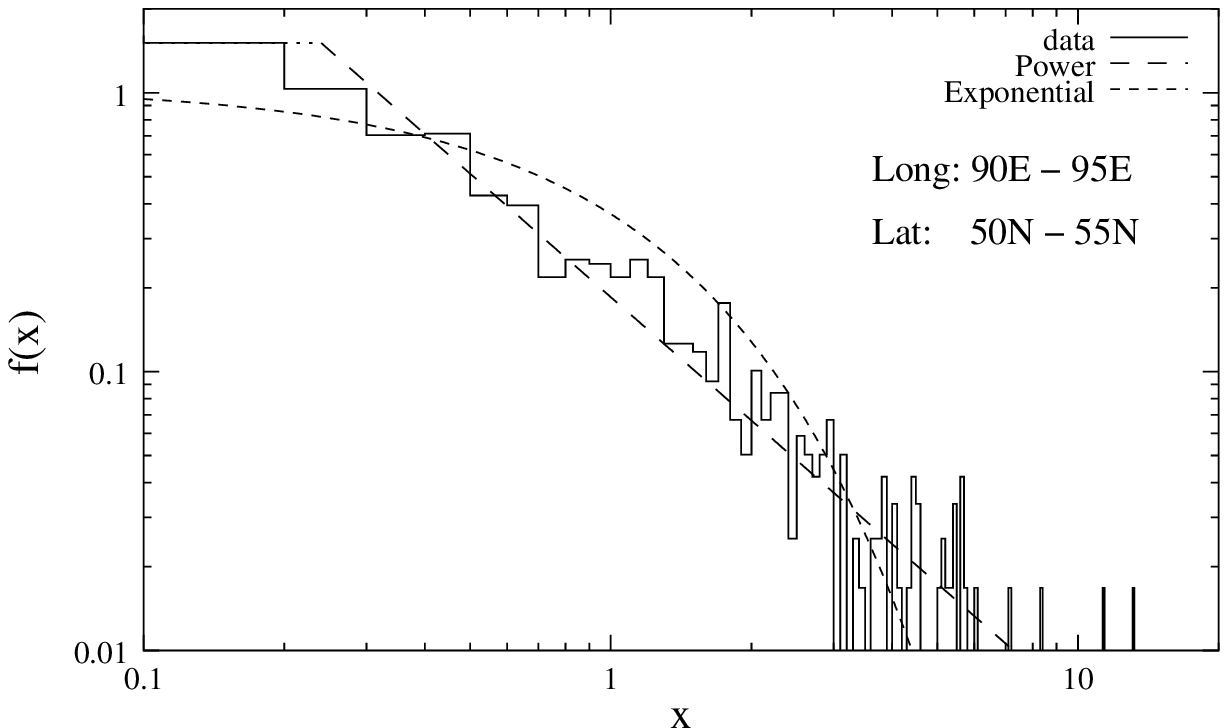}
\emi
}
\subf{
\bmi[t]{3in}
\includegraphics[scale=.55]{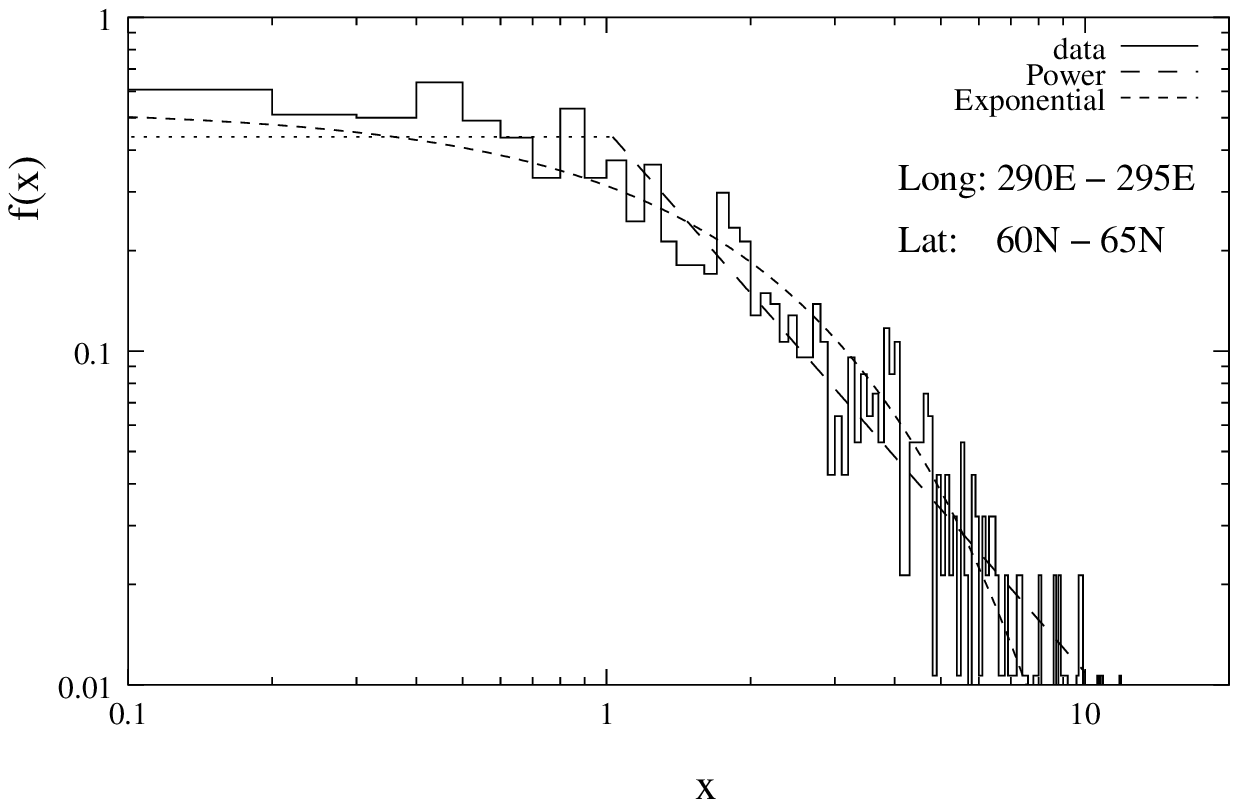}
\emi
}
\subf{
\bmi[t]{3in}
\includegraphics[scale=0.55]{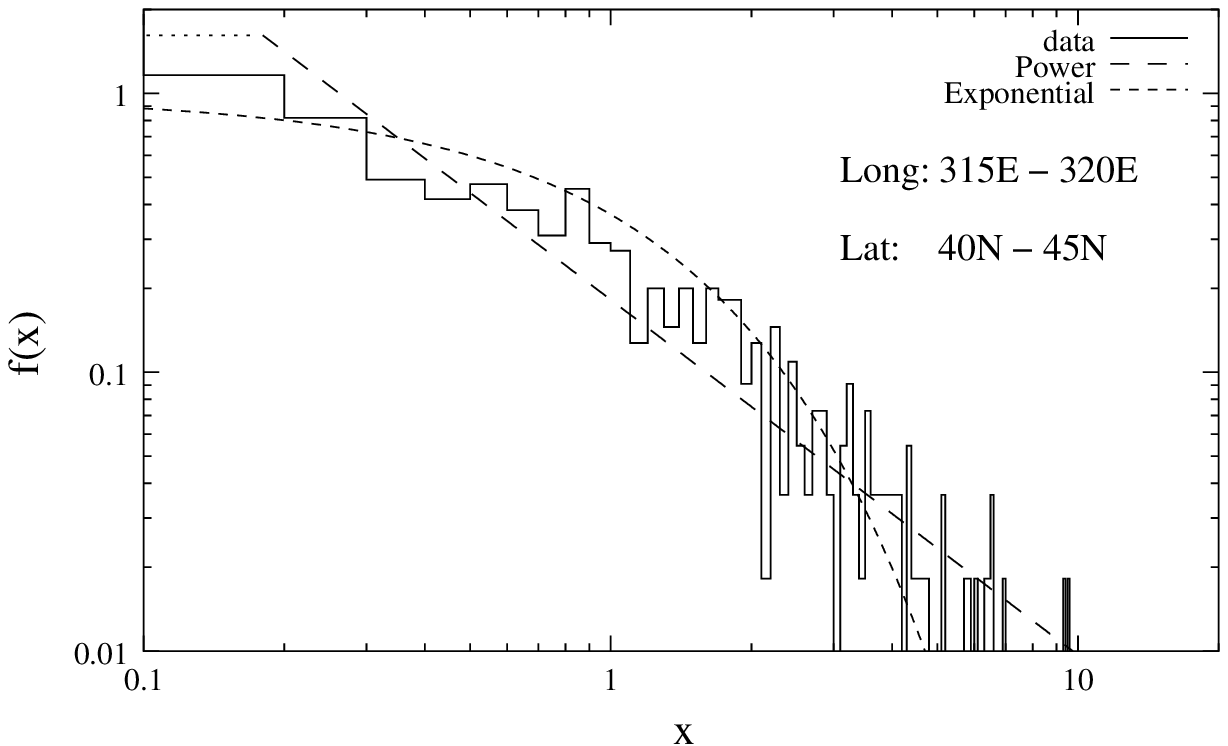}
\emi
}
\caption{Hourly rainfall distributions $f(x)$ of a sample of data sets  
at high latitudes. The variable $x$ is the rainfall rate in mm/hour.
The best fit power and exponential distributions
are also shown. The longitude and latitude range from where the 
data was taken is also indicated in each graph}
\label{High_Lat}
\efi

\end{document}